\documentclass{llncs}
\usepackage[T1]{fontenc}
\usepackage[english]{babel}
\usepackage{graphicx}
\usepackage{microtype}
\usepackage{amsmath,amssymb}
\usepackage{booktabs}
\usepackage{multirow}
\usepackage{array}
\usepackage{xcolor}
\usepackage{listings}
\usepackage{tikz}
\usepackage{tabularx}
\usepackage{makecell}
\usepackage{float}
\usepackage{placeins}
\usepackage[hidelinks]{hyperref}
\usepackage{orcidlink}
\usetikzlibrary{positioning,arrows.meta,shapes.geometric,fit}

\lstdefinestyle{promptstyle}{
  basicstyle=\scriptsize\ttfamily,
  breaklines=true,
  breakatwhitespace=false,
  frame=single,
  framesep=3pt,
  framerule=0.3pt,
  rulecolor=\color{gray!50},
  xleftmargin=4pt,
  xrightmargin=4pt,
  aboveskip=2pt,
  belowskip=2pt,
}
\graphicspath{{figures/}}
\newcommand{\toolname}{RAG-NIMBF}
\title{RAG-NIMBF: A Non-Invasive, Modular Benchmarking Framework for Automated RAG Configuration Analysis}
\author{
  Niclas Hart\thanks{Corresponding author}$^{,}$\inst{1}\orcidlink{0009-0009-9590-2915}
  \and
  Nicolas Weeger\inst{1}\orcidlink{0009-0007-6342-0693}
}

\institute{
  Ansbach University of Applied Sciences\\
  \email{n.hart18293@hs-ansbach.de, nicolas.weeger@hs-ansbach.de}
}
\begin{document}
\raggedbottom
\maketitle
\begin{abstract}
Retrieval-Augmented Generation (RAG) has become the dominant paradigm for grounding the outputs of large language models (LLM) in external knowledge. However, practitioners are faced with a vast configuration space comprising chunking, embedding, retrieval, reranking, prompting and generation models, and there is limited guidance on how these components interact. We present \toolname{}, a modular benchmarking framework that attaches to existing RAG pipelines as a non-invasive evaluation layer. It evaluates the configuration space through controlled ablation with end-to-end experiment tracking and is publicly available via GitHub.\footnote{\url{https://github.com/aiengineeringblueprints/RAG-NIMBF}} The evaluation combines LLM-based answer quality scores, retrieval correctness measures, and runtime indicators. This study validates the framework across multiple generation models on the SQuAD dataset. This process reveals non-obvious interactions and demonstrates how the experiments are tracked. The findings indicate that the retrieval balance of relevance and diversity substantially enhances the relevance and coverage of retrieved context in comparison to a conventional similarity search, with MMR raising nDCG@5 by +4 to +11 percentage points across the swept models. More detailed prompt instructions meaningfully increase how faithfully generated answers stick to the retrieved evidence, especially for capable and reasoning models (+10 to +13pp on Qwen3.6-27B, GPT-4o-mini and Qwen3.5-think, but negligible on the 20B generator). Larger generation models improve answer quality at the cost of higher latency, yet the local Qwen3.6-27B matches GPT-4o-mini at mean faithfulness 0.89, and the frontier reference does not dominate the smaller generators on RAG-relevant metrics. Overall, \toolname{} transforms the configuration space of RAG into reproducible, comparable experiments. By attaching to existing production pipelines in a non-invasive way, it enables practitioners to systematically evaluate and improve deployed systems without disrupting existing workflows. This replaces ad hoc choices with evidence-based decisions.

\end{abstract}

\section{Introduction}
\label{sec:intro}

Retrieval-Augmented Generation (RAG)~\cite{lewis2020rag} has emerged as the standard architecture for building knowledge-grounded question-answering systems. Rather than relying solely on parametric knowledge stored in model weights, RAG pipelines retrieve relevant documents from an external corpus and condition answer generation on the retrieved context. This approach reduces hallucination~\cite{shuster2021retrieval} and enables domain-specific applications without costly fine-tuning.

A production RAG pipeline involves multiple interacting components:
\begin{enumerate}
  \item A chunking strategy splits source documents into retrievable segments.
  \item An embedding model maps chunks and queries to a shared vector space.
  \item A retrieval method selects the most relevant chunks.
  \item An optional reranker re-orders them.
  \item A prompt template frames the generation task.
  \item A generation model produces the final answer~\cite{gao2024survey}.
\end{enumerate}
Each component offers multiple design choices resulting in a large combinatorial configuration space. Even small changes in chunk size, retrieval strategy, or prompt wording can shift faithfulness scores by more than ten percentage points~\cite{chen2024benchmark}. However, practitioners often select configurations based on intuition, vendor claims, or community defaults rather than systematic evaluations against their own data~\cite{es2023ragas,saadfalcon2024ares}, and such defaults are highly sensitive to corpus and model~\cite{ammar2025optimizing,ray2024ragserve}. For smaller organisations, building RAG on consumer-grade hardware rather than dedicated GPU clusters is the only economically viable path~\cite{weeger2025blueprints}, so this study covers models that run on such hardware. Although several evaluation frameworks exist~\cite{es2023ragas,saadfalcon2024ares,ragchecker2024}, their primary focus is on scoring a given pipeline rather than treating the configuration itself as the object of study. Systematic sweeps, component-interaction analysis, and reproducibility tracking are typically out of scope, and none of them are compatible with existing RAG pipelines.

These issues are addressed by \toolname{} (RAG-Non-Invasive Modular Benchmarking Framework), a framework which is designed around three core principles:
\begin{enumerate}
  \item Non-invasive integration: The framework wraps existing RAG components without modifying model or retriever internals. It connects to Ollama and OpenAI compatible endpoints (vLLM, TGI, \ldots) through provider routing, requiring only model identifier strings and access data.
  \item Systematic configuration exploration: A declarative environment configuration specifies parameter ranges for all components. The framework generates the cartesian product and benchmarks every combination, producing and tracking structured results suitable for direct comparison.
  \item Hybrid evaluation: Each run is assessed through a combination of answer-quality, retrieval, generation, and runtime metrics, so that configuration effects can be inspected from multiple complementary angles.
\end{enumerate}

\toolname{} is validated by benchmarking several configurations across different generation models spanning local, cloud-hosted, and API deployments on a 100-question sample from the SQuAD dataset~\cite{rajpurkar2016squad}. Experiment tracking is built into the framework itself, with every run automatically logged in MLflow. This includes parameters, metrics, artifacts and traces, so the results reported here are fully reproducible without additional bookkeeping.

The following design choices materialise in three concrete contributions:
\begin{enumerate}
  \item A modular, non-invasive benchmarking framework for systematic RAG configuration analysis, with integrated MLflow tracking that makes every sweep reproducible by default.
  \item A hybrid evaluation methodology combining answer-quality, retrieval, and generation metrics with runtime measurements.
  \item An empirical study revealing non-obvious interactions between chunking, retrieval, prompting, and generation components, with explicit focus on models that fit on consumer-grade hardware so that smaller organisations can adopt the findings without dedicated AI infrastructure.
\end{enumerate}

\section{Related Work}
\label{sec:related}

Lewis et al.~\cite{lewis2020rag} introduced RAG by combining dense passage retrieval~\cite{karpukhin2020dpr} with sequence-to-sequence generation. Subsequent systems have extended retrieval into pre-training, self-reflection, dynamic retrieval, graph-based reasoning and evidence checking~\cite{guu2020realm,asai2024selfrag,jiang2023active,edge2024graphrag,yu2024chainofnote}. Although surveys~\cite{gao2024survey} demonstrate that output quality is affected by chunking, retrieval, and prompt design, they offer no tooling for the systematic comparison of configurations.

Several frameworks evaluate a single pipeline configuration. RAGAS~\cite{es2023ragas} introduced LLM-based metrics (faithfulness, answer relevance, context precision/recall). ARES~\cite{saadfalcon2024ares} uses prediction-powered inference without gold references. 
\newline
RAGChecker~\cite{ragchecker2024} decomposes performance into retrieval and generation diagnostics. Chen et al.~\cite{chen2024benchmark} benchmark LLMs on RAG but evaluate only model-level differences. Observability platforms such as LangSmith~\cite{langsmith2024} and Langfuse~\cite{langfuse2024} provide tracing and cost telemetry without Cartesian-product sweeps. However, none of these systems sweep the multi-component configuration space systematically.

A second class provides end-to-end RAG frameworks. LangChain~\cite{chase2022langchain} focuses on application composition rather than evaluation. FlashRAG~\cite{jin2024flashrag} provides a modular toolkit comprising retrievers, generators, datasets, and a benchmark runner over preset configurations. However, it adopts an opinionated in-framework pipeline, meaning components must conform to its abstractions. Furthermore, evaluation couples neither IR, NLG, nor runtime indicators under a single non-invasive layer. DSPy~\cite{khattab2023dspy} optimises prompt/few-shot choices within a configured pipeline rather than enumerating the Cartesian space. CRUD-RAG~\cite{lu2024crudrag} contributes a Chinese-language benchmark over fixed pipelines. 
\newline
BERGEN~\cite{rau2024bergen} benchmarks retrieval and reader modules under fixed generation pipelines, but does not consider chunking, prompts, or runtime. RAGPerf~\cite{li2026ragperf} profiles production-like latency and throughput but does not enumerate the configuration space or couple IR/NLG with runtime.

The individual components have been studied in isolation. The lost-in-the-middle effect~\cite{liu2024lost}, MMR~\cite{carbonell1998mmr}, MTEB~\cite{muennighoff2022mteb}, and Sentence-BERT~\cite{reimers2019sentence}. These studies examine one component at a time, leaving interaction effects unexplored. Similarly, RAG benchmarks are usually dataset-centric, such as SQuAD~\cite{rajpurkar2016squad}, rather than configuration-centric.

By contrast, \toolname{} is designed as a non-invasive evaluation layer that wraps existing components using provider routing. It treats every pipeline stage as a swept dimension and fuses RAGAS-based quality, IR, NLG, and runtime metrics under MLflow-tracked runs, leaving the host pipeline intact.

\section{Framework Architecture}
\label{sec:methodology}

\toolname{} treats the configuration of a RAG pipeline, not the pipeline itself, as the object of study. Given a declarative specification of parameter ranges, it runs the cartesian product of all combinations and evaluates each run with a hybrid metric suite, leaving the host pipeline intact.

The design was guided by three principles:
\begin{enumerate}
  \item Non-invasive integration: components are addressed through model-identifier strings and provider routing, so existing RAG setups attach without modifying model or retriever internals..
  \item Systematic configuration exploration: an exhaustive cartesian sweep instead of surrogate-based search, so interaction effects across stages remain observable.
  \item Hybrid evaluation: RAGAS, IR, NLG, and runtime metrics are logged jointly to MLflow, so no single score can be optimised in isolation.
\end{enumerate}

\begin{figure}[H]
\centering
\includegraphics[width=\textwidth]{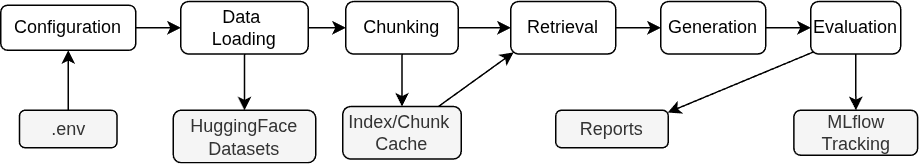}
\caption{Architecture of \toolname{}.}
\label{fig:architecture}
\end{figure}

\toolname{}'s pipeline proceeds through six stages, configuration, data loading, chunking, retrieval, generation, and evaluation, each implemented as an independent module. Figure~\ref{fig:architecture} shows the overall data flow, a declarative configuration drives every downstream stage, intermediate artifacts (chunks, vector indices, retrieved contexts) are cached for reuse across configurations, and all parameters and metrics are logged to MLflow.

\subsection{Configuration and Data Loading}
\label{subsec:config}

The framework reads all parameters from a single environment file. Users specify comma-separated lists for each sweepable parameter and the framework computes their cartesian product. The framework supports sweepable generation models, chunking strategies, chunk sizes, overlap, retrieval method, top-k, prompt templates, HyDE, and reranking.
We deliberately favour an exhaustive Cartesian sweep over more sample-efficient alternatives such as Bayesian optimisation or random search. Interaction effects across pipeline stages, for example chunk size against retrieval strategy or prompt template against model size, are the primary object of study. A surrogate-based optimiser would only reveal the best-performing point while leaving the interaction surface unobserved. The Cartesian product yields the full grid required for cross-configuration comparison and ablation. Model identifiers use provider:name syntax with automatic provider detection, e.g.\ ollama:qwen3:32b and openai:Qwen/Qwen3-32B-AWQ. A separate base URL can be configured per role, including generation LLM, critic LLM and embedding. This enables distributed setups and improves reproducibility by pinning each component to a fixed endpoint.

A registry-based adapter pattern maps dataset identifiers to loading functions. Each adapter converts a HuggingFace dataset into a unified \{question, ground\_truth, context\} format. Built-in adapters cover SQuAD~\cite{rajpurkar2016squad}, RagBench, T2-RagBench, Ragas-WikiQA, and RagPerf-Wikipedia-NQ. For datasets with a shared corpus, documents are loaded once and reused across configurations.

\subsection{Chunking}
\label{subsec:chunking}

The framework supports three chunking strategies via LangChain text splitters~\cite{wolf2020transformers}:
\begin{enumerate}
  \item Recursive splitting with a separator hierarchy and a robust default for prose.
  \item Fixed-size splitting using a token/character window with overlap and a deterministic baseline.
  \item Semantic splitting using embedding-similarity breakpoints to preserve topical coherence.
\end{enumerate}
Sentence-aware, markdown-aware, and late chunking were considered but not adopted due to high variance on noisy text, no structural markup in the corpus and compute-heavy long-context requirements respectively. For semantic chunking, chunk size and overlap are ignored since boundaries are determined by embedding similarity.

\subsection{Retrieval and Indexing}
\label{subsec:retrieval}

Chunks are embedded and stored in ChromaDB or LanceDB. We chose these two over FAISS, Milvus, Qdrant, Weaviate, and pgvector for their local-first persistence, low operational cost, and sufficient scale for our evaluation. Vector stores are cached using content fingerprint keys, allowing index reuse across configurations that share the same chunking and embedding setup.

Retrieval supports similarity search ($k$ nearest neighbours) and Maximal Marginal Relevance (MMR)~\cite{carbonell1998mmr}, which balances relevance with diversity through a tunable $\lambda$ parameter. An optional HyDE step generates a hypothetical answer from the query and uses it as the retrieval input. An optional cross-encoder reranker re-scores retrieved documents and selects the top-$n$ for generation. Sparse/hybrid retrieval (BM25, dense+sparse fusion) is out of scope.

\subsection{Generation and Provider Routing}
\label{subsec:generation}

The framework routes requests to Ollama or OpenAI-compatible endpoints based on the model prefix. A wrapper normalises response formats for RAGAS compatibility. Answer post-processing strips <think> traces emitted by reasoning models before scoring, so that faithfulness reflects the final answer rather than intermediate chain-of-thought, keeping reasoning and non-reasoning models comparable under the same evaluation protocol.

\subsection{Evaluation}
\label{subsec:evaluation}

Each configuration is evaluated through four complementary metric layers (Table~\ref{tab:metrics}).

\begin{table}[H]
\centering
\caption{Metric definitions per evaluation layer (RAGAS layer from Es et al.~\cite{es2023ragas}).}
\label{tab:metrics}
\scriptsize
\begin{tabular}{@{}lll@{}}
\toprule
Layer & Metric & Definition \\
\midrule
RAGAS   & Faithfulness        & $|\{c_i \in \mathcal{C}_{\text{ans}} : c_i \text{ entailed by } \mathcal{C}_{\text{ctx}}\}|/|\mathcal{C}_{\text{ans}}|$ \\
        & Context recall      & $|\text{GT attr.\ to } \mathcal{C}_{\text{ctx}}|/|\text{GT}|$ \\
        & Semantic sim.       & $\cos(\text{emb}(\text{ans}), \text{emb}(\text{GT}))$ \\
\midrule
IR      & Hit@$k$             & $\mathbb{1}[|\mathcal{R}_k \cap \mathcal{G}| \geq 1]$ \\
        & Recall@$k$          & $|\mathcal{R}_k \cap \mathcal{G}|/|\mathcal{G}|$ \\
        & nDCG@$k$~\cite{jarvelin2002ndcg} & $\text{DCG}@k/\text{IDCG}@k$, $\text{DCG}@k = \sum_{i=1}^{k} \text{rel}_i / \log_2(i+1)$ \\
\midrule
NLG     & \makecell[lt]{BERTScore-F1, METEOR,\\ROUGE-L, BLEU~\cite{zhang2020bertscore,banerjee2005meteor,lin2004rouge,papineni2002bleu}} & surface-overlap vs.\ reference \\
\midrule
Runtime & TTFT, TPS           & time-to-first-token, tokens-per-second from server \\
\bottomrule
\end{tabular}
\end{table}

Here, $\mathcal{C}_{\text{ans}}$ denotes atomic claims in the answer, $\mathcal{C}_{\text{ctx}}$ the retrieved context, $\mathcal{R}_k$ the top-$k$ retrieved items, $\mathcal{G}$ the gold items, and $\text{rel}_i$ the relevance grade of item $i$. The RAGAS layer is computed by a separate critic LLM (Qwen3.5-397B-A17B, non-thinking mode) and measures answer quality against the retrieved context. The IR layer measures the quality of the retrieval against the gold passages in SQuAD and is therefore independent of the critic. The NLG layer measures surface-form similarity against the reference answer. The runtime layer records latency and throughput directly from the inference server. Reading the four layers jointly prevents any single metric from being gamed.

All parameters, metrics, and artifacts are logged to MLflow with a reproducibility bundle capturing the full environment state.

\section{Experimental Setup}
\label{sec:setup}

\subsection{Models}
\label{subsec:models}

The experimental matrix pairs single-workstation local generators with two frontier API references so that configuration effects can be observed within a single-GPU budget and against an upper bound. Table~\ref{tab:models} lists the six generation models, spanning dense transformers, mixture-of-experts (MoE) architectures, reasoning variants, and local plus API deployments. The local lineup targets the prosumer / single-accelerator tier that smaller organisations can realistically deploy without dedicated GPU clusters. The cloud APIs serve as upper-bound reference points.

\begin{table}[H]
\centering
\caption{Generation models with parameter counts.}
\label{tab:models}
\small
\begin{tabular}{lll}
\toprule
\textbf{Model} & \textbf{Params} & \textbf{Deployment} \\
\midrule
Qwen3-32B-AWQ\textsuperscript{1}        & 32B dense           & Workstation \\
GPT-OSS-20B\textsuperscript{2}          & 20B (3.6B active)   & DGX Spark (GB10) \\
Qwen3.5-35B-A3B\textsuperscript{3}        & 35B (3B active)     & GPU-Cluster \\
Qwen3.6-27B-NVFP4\textsuperscript{4}    & 27B dense           & Workstation \\
GPT-4o-mini\textsuperscript{5}          & $\sim$20-30B (est.) & GPU-Cluster \\
DeepSeek-V4-Pro\textsuperscript{6}      & 862B                & API \\
\bottomrule
\end{tabular}
\\[2pt]
\scriptsize
\textsuperscript{1}Huggingface, \url{https://huggingface.co/Qwen/Qwen3-32B-AWQ}, accessed: Jul.\ 20, 2026\\
\textsuperscript{2}Huggingface, \url{https://huggingface.co/openai/gpt-oss-20b}, accessed: Jul.\ 20, 2026\\
\textsuperscript{3}Huggingface, \url{https://huggingface.co/Qwen/Qwen3.5-35B-A3B}, accessed: Jul.\ 20, 2026\\
\textsuperscript{4}Huggingface, \url{https://huggingface.co/sakamakismile/Qwen3.6-27B-Text-NVFP4-MTP}, accessed: Jul.\ 20, 2026\\
\textsuperscript{5}OpenAI, \url{https://platform.openai.com/docs/models/gpt-4o-mini}, accessed: Jul.\ 20, 2026\\
\textsuperscript{6}Huggingface, \url{https://huggingface.co/deepseek-ai/DeepSeek-V4-Pro}, accessed: Jul.\ 20, 2026\\
\end{table}

The largest open-weight model available to us, Qwen3.5-397B-A17B, an MoE with 397B total and 17B active parameters, is deliberately excluded from the generation sweep and reserved for the RAGAS critic role. Using the strongest available open model as the judge keeps evaluation quality uniform across configurations and avoids confounding generator quality with judge weakness. Running the critic on the same H100 cluster in non-thinking mode holds the evaluation harness fixed while the generators vary. DeepSeek-V4-Pro is included as a frontier reference but supports only similarity retrieval over its API, so it appears in the corresponding subset of the configuration space only.

\subsection{Embedding and Dataset}
\label{subsec:embedding}

All configurations use nomic-embed-text:latest~\cite{nomic2024embed} (768-dim) as a fixed embedding backbone, an open, locally deployable model suitable for single-workstation benchmarking without external API cost. Holding the embedding model constant isolates the effects of chunking, retrieval strategy, prompt template, and generation model; embedding-model comparison is therefore out of scope.

We use SQuAD~\cite{rajpurkar2016squad} because it is a widely used extractive question-answering benchmark with gold passage supervision. This makes it possible to compute retrieval metrics (Hit@$k$, nDCG@$k$, Recall@$k$) against a critic-independent ground truth while also evaluating generated answers against short reference answers. Its public availability and broad use in QA and RAG studies make the results easier to reproduce and compare to prior work.

\subsection{Configuration Space}
\label{subsec:configspace}

Table~\ref{tab:config-space} summarizes the parameters swept in our experiments. The full grid contains $M \times C \times S \times O \times R \times T = 6 \times 2 \times 2 \times 2 \times 2 \times 2 = 192$ configurations. The empirical study reported here instantiates a deliberately smaller subset of that surface, which turns the experiment into a controlled ablation rather than a benchmark over every framework option.

\begin{table}[H]
\centering
\caption{Configuration space.}
\label{tab:config-space}
\small
\begin{tabular}{ll}
\toprule
\textbf{Parameter} & \textbf{Values} \\
\midrule
LLM Models ($M = 6$)              & See Table~\ref{tab:models} \\
Chunking Strategies ($C = 2$)     & Recursive, Semantic \\
Chunk Sizes ($S = 2$)             & 500, 1000 \\
Chunk Overlaps ($O = 2$)          & 100, 200 \\
Retrieval Methods ($R = 2$)       & Similarity, MMR \\
Prompt Templates ($T = 2$)        & Concise, Detailed \\
Retrieval Top-$k$                 & 12 \\
Reranker / HyDE                   & Disabled \\
Critic model (outside the sweep)  & Qwen3.5-397B-A17B (non-thinking) \\
\bottomrule
\end{tabular}
\end{table}

\subsection{Prompt Templates}
\label{subsec:prompts}

Figure~\ref{fig:prompt-templates} shows the two prompt templates used in the experiments. The Concise template enforces a single-sentence factual answer with few-shot exemplars. The Detailed template removes the exemplars and asks for a full-sentence answer with supporting reasoning.

\begin{figure}[H]
\centering
\begin{minipage}{0.48\textwidth}
\small Concise\\[2pt]
\begin{lstlisting}[style=promptstyle]
Answer the question using ONLY the
provided context.
RULES:
- Short, direct, factual (one
  sentence).
- No reasoning, no markdown.
- If absent: 'I cannot answer from
  the provided context.'
Examples:
Q: Who wrote the music for Star Wars?
A: John Williams.
\end{lstlisting}
\end{minipage}\hfill
\begin{minipage}{0.48\textwidth}
\small Detailed\\[2pt]
\begin{lstlisting}[style=promptstyle]
Answer the question based only on
the provided context.
Provide a clear, complete answer
in full sentences.
Include relevant calculations or
reasoning when applicable. Do not
start with 'Based on the provided
context'. Answer directly.

Human turn (both):
Context: {context}
Question: {question}
Answer:
\end{lstlisting}
\end{minipage}
\caption{The two prompt templates.}
\label{fig:prompt-templates}
\end{figure}

\subsection{Evaluation Protocol}
\label{subsec:evalprotocol}

RAGAS metrics (faithfulness, context recall, semantic similarity) are computed using Qwen3.5-397B-A17B as the critic model with a maximum of 4096 critic tokens per evaluation call. IR metrics (Hit@$k$, nDCG@$k$, Recall@$k$ for $k \in \{1, 3, 5\}$) are computed against gold document labels provided by the SQuAD dataset. NLG metrics use RoBERTa-large for BERTScore~\cite{zhang2020bertscore} and standard implementations for METEOR~\cite{banerjee2005meteor}, ROUGE-L~\cite{lin2004rouge}, and BLEU~\cite{papineni2002bleu}. All metrics are reported as means over the 100-question sample, with per-question values retained for the bootstrap and significance analyses reported in Section~\ref{sec:results}.

\section{Results}
\label{sec:results}

All metric values are means over 100 SQuAD questions per configuration. We organise findings into three research questions covering generation-model comparison, component effects and interactions, and runtime-quality tradeoffs. The results are preliminary and represent a subset of the planned evaluation.

\subsection{RQ1: How Do Generation Models Compare?}
\label{subsec:rq1}

Table~\ref{tab:model-comparison} aggregates results across all configurations for each model.

Three models share the top mean faithfulness at 0.89 (GPT-4o-mini, Qwen3.6-27B, and the reasoning-only Qwen3.5-think at 0.88), with Qwen3-32B and GPT-OSS-20B trailing at 0.83 and 0.77. The best single configuration reaches up to 0.96 faithfulness (GPT-4o-mini), and 0.95 is attained by Qwen3.5-think and Qwen3.6-27B, showing that strong grounding is attainable across architectures and parameter budgets. DeepSeek-V4-Pro leads METEOR at 0.56 but is restricted to similarity retrieval and therefore appears in the corresponding subset of the configuration space only. Among the local models, Qwen3.6-27B matches GPT-4o-mini on mean faithfulness (0.89). The faithfulness ranking is robust within $\pm 0.01$ based on a 95\% bootstrap CI over 2000 resamples.

Retrieval metrics are reported separately because they are generator-invariant by construction. Hit@$k$, nDCG@$k$, and Recall@$k$ are pure functions of the retrieved chunk set and the gold document, neither of which depends on the generator when the embedding model and retrieval configuration are held fixed. At fixed (\textit{chunking}, \textit{retrieval}, \textit{top-k}) cells, nDCG@5 varies by at most 0.07 across generators, driven by per-run question-subset differences. Aggregated across all runs, the global retrieval quality is nDCG@5 = 0.75, Hit@5 = 0.86, Recall@5 = 0.69. Per-configuration values are reported in Section~\ref{subsec:rq2}.

\begin{table}[H]
\centering
\caption{Model comparison: mean and (best) across all configurations. Retrieval metrics omitted from per-model rows because generator-invariant by construction; global retrieval quality reported in the final row.}
\label{tab:model-comparison}
\small
\begin{tabular}{lccc}
\toprule
\textbf{Model} & \textbf{Faith.} $\uparrow$ & \textbf{BS-F1} $\uparrow$ & \textbf{METEOR} $\uparrow$ \\
\midrule
Qwen3.5-think    & 0.88 (0.95) & 0.86 (0.93) & 0.49 (0.60) \\
GPT-4o-mini      & 0.89 (0.96) & 0.87 (0.93) & 0.49 (0.59) \\
DeepSeek-V4-Pro  & 0.88 (0.94) & 0.91 (0.92) & 0.56 (0.66) \\
Qwen3.6-27B      & 0.89 (0.95) & 0.89 (0.95) & 0.51 (0.61) \\
Qwen3-32B        & 0.83 (0.92) & 0.91 (0.97) & 0.48 (0.60) \\
GPT-OSS-20B      & 0.77 (0.84) & 0.90 (0.96) & 0.46 (0.60) \\
\midrule
\multicolumn{4}{l}{\textit{Retrieval quality (all 222 non-critic runs, generator-invariant)}} \\
nDCG@5 / Hit@5 / Recall@5 & 0.75 & 0.86 & 0.69 \\
\bottomrule
\end{tabular}
\end{table}

\subsection{RQ2: How Do Retrieval, Prompting, and Chunking Affect Quality?}
\label{subsec:rq2}

The retrieval and prompt comparisons below use the five-model subset with full coverage of the swept configuration space. Table~\ref{tab:retrieval} compares similarity search and MMR on this subset. MMR improves nDCG@5 by +4 to +11pp by reducing redundancy in the retrieved set, with the largest gains on GPT-4o-mini (+11pp, $0.68 \rightarrow 0.79$) and Qwen3.5-think (+10pp, $0.70 \rightarrow 0.80$), and the smallest on GPT-OSS-20B (+4pp). The faithfulness effect reverses sign with model size, the two smaller generators Qwen3-32B (+5pp) and GPT-OSS-20B (+6pp) gain faithfulness under MMR, while GPT-4o-mini ($-4$pp), Qwen3.5-think ($-2$pp) and Qwen3.6-27B ($-1$pp) lose some. The smaller generators benefit from the broader evidence mix that MMR provides, whereas the stronger generators already ground well on the most similar passage and are diluted by the additional context.

\begin{table}[H]
\centering
\caption{Retrieval strategy comparison: mean over chunk sizes, overlaps, and templates.}
\label{tab:retrieval}
\small
\begin{tabular}{llcccc}
\toprule
\textbf{Model} & \textbf{Strategy} & \textbf{Faith.} & \textbf{nDCG@5} & \textbf{BS-F1} & \textbf{METEOR} \\
\midrule
\multirow{2}{*}{Qwen3-32B}    & Sim. & 0.80 & 0.71 & 0.92 & 0.51 \\
                              & MMR  & 0.85 & 0.81 & 0.89 & 0.47 \\
\multirow{2}{*}{GPT-OSS-20B}  & Sim. & 0.74 & 0.71 & 0.92 & 0.46 \\
                              & MMR  & 0.80 & 0.75 & 0.89 & 0.45 \\
\multirow{2}{*}{GPT-4o-mini}  & Sim. & 0.90 & 0.68 & 0.88 & 0.50 \\
                              & MMR  & 0.86 & 0.79 & 0.82 & 0.47 \\
\multirow{2}{*}{Qwen3.6-27B}  & Sim. & 0.90 & 0.67 & 0.90 & 0.52 \\
                              & MMR  & 0.89 & 0.74 & 0.88 & 0.50 \\
\multirow{2}{*}{Qwen3.5-think} & Sim. & 0.88 & 0.70 & 0.85 & 0.47 \\
                               & MMR  & 0.86 & 0.80 & 0.82 & 0.47 \\
\bottomrule
\end{tabular}
\end{table}

Table~\ref{tab:template} shows the Detailed template raises faithfulness by +10 to +13pp on the cloud/large generators and the reasoning model (Qwen3.5-think +13pp, GPT-4o-mini +11pp, Qwen3.6-27B +10pp), while Qwen3-32B gains +7pp and GPT-OSS-20B essentially does not move (+1pp). The faithfulness gain trades off against METEOR (10-22pp drop with Detailed) because detailed, evidence-grounded answers diverge from the brief SQuAD references. BERTScore F1 is largely unaffected (0-4pp), confirming semantic stability. This lexical-vs-semantic divergence illustrates why multi-metric evaluation is necessary. Optimising METEOR alone would misleadingly favour under-specified answers.

Four combinations, namely 500/100, 500/200, 1000/100, and 1000/200, were tested on Qwen3-32B with recursive chunking. Effects are modest, faithfulness spans 0.70-0.87 and nDCG@5 spans 0.73-0.77 across the four settings. Smaller chunks of 500 characters yield slightly higher nDCG@5 (mean 0.76) than 1000-character chunks (mean 0.74) via finer granularity, while larger chunks marginally raise faithfulness at overlap 200 (0.87) via broader grounding. Semantic chunking yields identical results across all size/overlap combinations, since boundaries follow embedding-similarity breakpoints rather than character counts.

\begin{table}[H]
\centering
\caption{Prompt template comparison: mean over all chunk sizes, overlaps, and retrieval strategies.}
\label{tab:template}
\small
\begin{tabular}{llcccc}
\toprule
\textbf{Model} & \textbf{Template} & \textbf{Faith.} & \textbf{nDCG@5} & \textbf{BS-F1} & \textbf{METEOR} \\
\midrule
\multirow{2}{*}{Qwen3-32B}    & Concise  & 0.80 & 0.74 & 0.89 & 0.58 \\
                              & Detailed & 0.87 & 0.74 & 0.91 & 0.41 \\
\multirow{2}{*}{GPT-OSS-20B}  & Concise  & 0.76 & 0.74 & 0.91 & 0.56 \\
                              & Detailed & 0.77 & 0.74 & 0.91 & 0.34 \\
\multirow{2}{*}{GPT-4o-mini}  & Concise  & 0.83 & 0.75 & 0.83 & 0.53 \\
                              & Detailed & 0.94 & 0.75 & 0.85 & 0.42 \\
\multirow{2}{*}{Qwen3.6-27B}  & Concise  & 0.84 & 0.76 & 0.84 & 0.55 \\
                              & Detailed & 0.94 & 0.75 & 0.88 & 0.43 \\
\multirow{2}{*}{Qwen3.5-think} & Concise  & 0.81 & 0.74 & 0.81 & 0.52 \\
                               & Detailed & 0.94 & 0.75 & 0.85 & 0.42 \\
\bottomrule
\end{tabular}
\end{table}

\subsection{RQ3: Performance--Quality Tradeoffs}
\label{subsec:rq3}
Table~\ref{tab:perf-quality} reports the raw inference measurements for 15 prompts and a maximum of 1024 output tokens, for both sequential and parallel processing. DeepSeek-V4-Pro is excluded because its third-party-API latency is dominated by network and is therefore not comparable to the other checkpoints on identical hardware.

\begin{figure}[H]
\centering
\includegraphics[width=\textwidth]{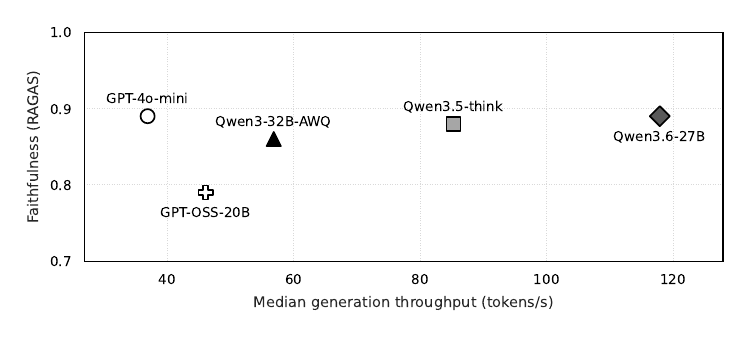}
\caption{Faithfulness vs.\ median generation throughput.}
\label{fig:pareto}
\end{figure}

Qwen3.6-27B-NVFP4 leads sequential throughput (127.5 TPS) and parallel throughput among the locals (70.5 TPS). Figure~\ref{fig:pareto} plots throughput against faithfulness. Qwen3.6-27B-NVFP4 anchors the high-throughput end, GPT-4o-mini and Qwen3.5-think the high-quality end. GPT-OSS-20B sits off the quality-speed trade-off.

\begin{table}[H]
\centering
\caption{Runtime measurements (N=15, max 1024 output tokens). Seq.=sequential, Par.=parallel, TPS=tokens per second, TTFT=time-to-first-token.}
\label{tab:perf-quality}
\small
\setlength{\tabcolsep}{3pt}
\begin{tabular}{lccccc}
\toprule
\textbf{Model} & \textbf{Seq.\,t (s)} & \textbf{Par.\,t (s)} & \textbf{Seq.\,TPS} & \textbf{Par.\,TPS} & \textbf{TTFT (s)} \\
\midrule
GPT-OSS-20B        & 77.09 & 44.56 & 53.9  & 22.4 & 0.47 \\
Qwen3-32B-AWQ       & 87.13 & 22.23 & 58.9  & 45.4 & 0.30 \\
GPT-4o-mini        & 54.93 & 28.76 & 79.4  & 59.1 & 0.45 \\
Qwen3.5-think        & 80.19 & 26.43 & 78.8  & 68.5 & 3.50 \\
Qwen3.6-27B-NVFP4   & 38.59 & 24.80 & 127.5 & 70.5 & 1.86 \\
\bottomrule
\end{tabular}
\end{table}

\section{Discussion}
\label{sec:discussion}

The primary contribution of this work is \toolname{}, a non-invasive benchmarking layer for systematically evaluating configurations across the RAG configuration space. Empirical findings suggest that the framework can help identify meaningful effects across pipeline stages, offering insights into the behaviour of different RAG configurations on consumer-grade hardware.

\subsection{What the Framework Reveals}
\label{subsec:reveals}

As \toolname{} sweeps through every stage of the pipeline rather than fixing any component, the relative weight of each lever becomes apparent. The retrieval strategy and prompt wording can increase faithfulness and nDCG@5 by up to 11 percentage points, whereas the chunk size and overlap stay within 2-5 points. Interactions matter too, detailed prompts have almost no effect on GPT-OSS-20B but add +11 points on GPT-4o-mini. MMR reverses its faithfulness effect across model sizes.

The metrics also need to be considered together. Detailed prompts increase faithfulness but lower METEOR by 10-22 points, because SQuAD references are short and evidence-based answers differ in surface form while staying semantically aligned.

\subsection{Recommendations}
\label{subsec:recommendations}

At fixed compute, the levers fall out of the sweep in a clear order. Prompting is the lowest-integration lever on capable models. Retrieval comes next, as measured by MMR's nDCG@5 gains. Chunk size and overlap have smaller effects. Once these are set, the generator can be scaled appropriately. In our sweep, Qwen3.6-27B is a reasonable substitute for the tested cloud APIs.

\subsection{Limitations}
\label{subsec:limitations}

The sweep leaves reranking, HyDE, and embedding-model comparisons out, holds top-$k$ and the MMR $\lambda$ fixed, and is as of now only evaluated on the SQuAD dataset. SQuAD is short, extractive, mostly single-passage QA, results may not transfer to long-document, multi-hop, noisy, or enterprise corpora. All quality assessments are automated and therefore hold relative to the RAGAS critic. The prompt templates were authored once and applied uniformly. It is not yet clear if different model families benefit from prompt structures that are structurally different.

\section{Conclusion}
\label{sec:conclusion}

This paper presents \toolname{}, a non-invasive, modular benchmarking framework for systematic RAG configuration analysis. Across the swept SQuAD configurations, prompt specificity had the strongest effect on faithfulness, MMR improved retrieval coverage, and chunk-size effects were secondary. The results show why RAG configurations must be evaluated with multiple metrics: lexical scores, semantic similarity, retrieval quality, and runtime can move in different directions. Runtime measurements indicate that Qwen3.6-27B-NVFP4 can match GPT-4o-mini on mean faithfulness at lower latency, while the frontier reference model does not dominate on RAG-relevant metrics. Future work will extend the sweep to additional datasets, reranking, HyDE, and embedding models, and align \toolname{} with on-premises enterprise RAG architectures such as the Weeger et al.\ blueprint~\cite{weeger2026ragblueprint}.

\section{Disclosure of Interests}
The authors have no competing interests to declare that are relevant to the content of this article.

\bibliographystyle{splncs04}
\bibliography{references}
\end{document}